# An individual-level assessment of the relationship between spin-off activities and research performance in universities[1]

**Abstract**

One of the most problematic aspects in the creation of spin-offs by university personnel concerns the relationship between entrepreneurial activity and research activity by researcher-entrepreneurs. The literature has expressed varying and opposing views as to the nature of the relationship but very little has been produced to empirically legitimate one position or another. The present work proposes to address this shortcoming by exploring the relationship existing between academic spin-off generation and the research performance of enterprise founders. The study investigates whether, and to what extent, scientific performance by academic entrepreneurs is different than that of their colleagues, and if the involvement in entrepreneurial activity has an influence on the individual's research activity. The research questions are answered by considering all spin-offs generated by Italian universities over the period 2001-2008 and evaluating, through a bibliometric approach, the scientific performance of founders relative to that of their colleagues who carry out research in the same field. The data show better scientific performance by the researcher-entrepreneurs than that of their colleagues and in addition, although there are some variations across fields, the creation of a spin-off does not seem, on average, to have negative effects on the scientific performance of the founders.

## 1. Introduction

Since the approval of the United States Bayh-Dole Act, in the 1980, there has been a multiplication of studies concerning the so-called third mission of universities, which are now called to contribute directly to economic development, through collaboration with industry and exploitation of research results (Etzkowitz 2003). Among the means that universities can adopt to pursue this mission, increasing attention is being given to the phenomenon of research spin-offs. Among the various forms of academic entrepreneurship (patents, awarding of licenses, cooperation contracts with industry, spin-offs), the founding of new technology-based ventures is, indeed, one of the most direct and effective ways in which new knowledge and technology is commercialized (Davenport et al. 2002). Spin-off companies tend to locate close to their originating institutions and then become valuable entities for the local economic development and for the economies of agglomeration (Zucker et al. 1998). As well, university spin-offs create jobs for highly skilled graduates and show strong economic effects for regional communities (Rothaermel and Thursby 2005; Shane 2004a), since they serve as valuable sources of knowledge spillover for other companies (Benneworth and Charles 2005).

The belief in the importance of academic spin-off for economic development explains the increasing diffusion of governments' interventions aimed at fostering such form of entrepreneurship, as well as the occurrence of studies seeking to better understand and address the drivers that shape spin-off activity in higher education institutions (Chang et al. 2009; O'Shea et al. 2007; Wright et al. 2006; Murray 2004; Shane 2004a, 2004b; Di Gregorio and Shane 2003).

In addition to these mentioned studies, another stream of literature underlines the negative effects that this new form of university behavior can have on the traditional role of public research institutions, arguing that involvement in entrepreneurial activities can negatively impact on a researcher's scientific outcome (Nelson 2001; Metcalfe 1998; Feller 1990; Nelson 1959). The argument is based on the idea that there are different cultures, attitudes and incentive systems in academia and the private sector, particularly with respect to disclosure versus secrecy of research output (Dasgupta and David 1994; 1987). One of the cornerstones of the academic ethic is, indeed, the publication of research results and the opportunity for open discussions among colleagues. Companies, on the other hand, have needs and responsibilities to protect the value of their investments. The differences in public and private research incentive systems is held to create challenges to the academic entrepreneurs' scientific performance, concerning the types of possible research activity, dissemination of information, and access to research results (Fabrizio and Di Minin 2008; Azoulay et al. 2006; Jacobsson 2002; Florida and Cohen 1999; Hane 1999).

These concerns have spawned a number of empirical studies that investigate the broader implications of the increasing involvement of academics in entrepreneurial activities – such as patenting, consulting, collaboration with companies, spin-offs - on their scientific productivity.

In the context of these studies, the aim of our paper is to specifically examine the relationship between spin-off generation and the research performance of enterprise founders. This aspect has been little examined by the literature, which has instead concentrated on analysis of the relation between academic scientific productivity and other forms of entrepreneurship, such as patent protection. Further, the few studies which have actually examined the relation between spin-off creation and scientific productivity of the academic entrepreneurs have arrived at contrasting results. In particular, those examining the drivers for spin-off creation have shown that, at the overall level of the university, there is a positive relation between scientific excellence of faculty, and number of spin-offs achieved by the university (Van Looy et al. 2011; O'Shea et al. 2008; Landry et al. 2006; Powers and McDouglas 2005; Di Gregorio and Shane 2003). On the other hand, studies examining the same relationship at the individual level, have shown contrasting results (Lowe and Gonzales-Brambila 2007; Buenstorf, 2009).

Moreover these latter studies have generally tested their hypotheses by analyzing entrepreneurial and research activities of academics belonging to only a few scientific disciplines or based on few highly reputed research institutions. No relevant studies have yet analyzed the relationship between researchers' spin-off involvement and scientific performance by considering, at the same time: i) researchers belonging to many different disciplines and ii) working for large numbers of large and influential, and smaller and less famous research institutions and universities.

The current work is intended to assist in addressing the above gaps in knowledge by exploring and identifying the relationships between the spin-off generation and the research performance of enterprise founders. The paper tries to answer two specific research questions:

i.  *Do faculty members who found spin-off ventures have higher research performance than their colleagues (i.e. publish more and higher impact papers)?*

ii. *Does research performance of entrepreneurial academics, contrasted to that of their*

*colleagues, decrease after the founding of spin-off ventures?*

In trying to respond to these questions, the study considers all the spin-off generated by Italian universities in the period 2001-2008, evaluates the scientific performance of their respective founders, and compares it with that of all their national colleagues belonging to the same scientific discipline.

The paper is articulated as follows: the next section presents the theoretical background underlying the work; Section 3 presents the Italian university context with regard to spin-off laws and policies; Section 4 describes the method for constructing the dataset and the methodological choices involved in the bibliometric approach; Section 5 answers the research questions through presentation of the results from the elaborations, while the final section presents a summary of the work and some considerations on possible policy implications.

## 2. Theoretical background: faculty entrepreneurship and individual research productivity

Starting from Florida and Cohen's (1999) position, many scholars have investigated the relationship between a university scientist's entrepreneurial activity and his/her research performance, but there are only a few contributions on the specific relationship between spin-off generation and research performance. Entrepreneurial activity by a university researcher can, indeed, encompass a large variety of non-traditional behaviors, such as seeking patent protection, provision of consulting, engagement in contracts with private companies, and forming new companies through spin-off processes (Debackere and Veugelers 2005; Louis et al. 1989).
Although all these forms of faculty entrepreneurship have the potential to enhance or detract from academic entrepreneurs' productivity levels, most studies have been exclusively focused on examining the relationship between patent and publication intensity. Whether and how publication intensity is related to patent intensity is quite controversial though. Some scholars who have empirically examined this latter relationship have verified, contrary to the assertions of Florida and Cohen (1999), the existence of a positive relationship between patenting and publication outcomes of university researchers, both in terms of publication numbers (Carayol 2007; Czarnitzki et al. 2007; Stephan et al. 2007; Breschi et al. 2006; Meyer 2006; Van Looy et al. 2006; Lach and Shankerman 2003; Agrawal and Henderson 2002; Adams and Griliches 1998) and in terms of publication quality (Breschi et al. 2007; Czarnitzki et al. 2007; Azoulay et al. 2006; Van Looy et al.

2006). These results are consistent with the works of Zucker et al. (2002) and West (2008) who suggest that university-to-firm technology transfers involving breakthrough invention in biotechnology typically involve star scientists (see also Zucker and Darby 2001; 2007; Zucker et al. 1998).

On the contrary, other scholars have demonstrated that the relationship between patenting activity and publication performance is not always positively related. Fabrizio and Di Minin (2008) showed that the relationship depends on the number of patents filed. The authors found that when the number of patents increases, the positive relationship between researchers' publication and patenting activities declines, and the average number of citations to publications falls. Wong and Singh (2010), analyzing the relationship between patenting activities and scientific productivity of 281 leading universities world-wide, found different results for universities located in different geographical areas. In North American universities both the quantity and quality of scientific publications are positively related to patenting activities; in European and Australian/New Zealand universities, only quantity of publications is, while in universities outside North America and Europe/Australia/New Zealand, only quality of publications is.

Other authors have analyzed the effects on research productivity of other kinds of academic entrepreneurial behaviors, such as licensing and other forms of collaboration and technology transfer from academia to private firms. In particular Chang and Yang (2008) analyzing the scientific productivity of 229 Taiwanese academic inventors, found a significant difference between the inventors involved only in patenting activities and the inventors involved also in licensing activity. The former maintain high scientific productivity while the latter register a delay in publication activities and a relevant change in scientific involvement (from basic to applied research). Thursby and Kemp (2002) showed that the lower the research quality of a university the more efficient the university tends to be in commercial activity. They interpret this as being the result of greater specialization in basic research of the higher quality research faculty. Van Looy et al. (2004), instead, comparing the research performance of academics involved in private sector projects with that of their peers, found that involvement in contract research does not negatively affect the researcher's scientific outcomes. A similar study was conducted by Manjarrés-Henríquez et al. (2008). The authors, analyzing a sample of 2,135 researchers at two high-level Spanish research institutions, showed that when university-industry relationships concern low technological-scientific levels (technological support, consultancy, and similar activities) research performance suffers, but when university-industry relationships concern activities with high scientific-technological content (i.e. R&D contracts) the impact on research performance is positive. Abramo et al. (2009) analyzed the correlation between university research performance and intensity of collaboration with private companies, for science and engineering. They found a strong correlation in biology and, to a lesser degree, in physics and earth sciences. Toole and Czarnitzki (2010) explored the academic brain drain phenomenon, which occurs when academics take employment positions at for-profit firms. They found that the negative impact on knowledge production in the not-for-profit research sector is nontrivial.

Similarly other scholars (Gulbrandsen and Smeby 2005) found that faculty members who received research funds from industry published more articles than peers without industrial financial support. On the other hand, in a study on Italian universities Bonaccorsi et al. (2006) showed a tradeoff between research for publication and research for industrial use or patenting. They demonstrated that although collaboration with industry (as indicated by the average percentage of university budgets funded by industry from 1994 to 1999) might initially improve aggregate productivity, beyond a certain level it appeared to negatively affect publication profiles in some universities, possibly because of the difficulties in meeting the increasing expectations of industry as collaboration increases.

Finally, even if some studies conducted at university level suggest that the scientific eminence of university can be considered a driver of spin-off creation (Van Looy et al. 2011; O'Shea et al. 2008; Landry et al. 2006; Powers and McDouglas 2005; Di Gregorio and Shane 2003), very few examined the relationship between the most extreme expression of academic entrepreneurship – the generation of spin-off firms - and scientific performance at individual level. In particular only two empirical contributions have examined the effects of firm creation on the scientific performance of the academic founders (Larsen 2011). Investigations by Buenstorf (2009) included an examination of how founding a for-profit firm affects academic research productivity. He examined how four indicators of entrepreneurial behavior influenced the quantity and quality of publications of a sample of elite German scholars who held director positions at the Max Planck Institute, for the period 1985-2004. The four indicators considered were: i) disclosing inventions; ii) disclosing inventions that were licensed to the private sector; iii) disclosing inventions that were licensed to spin-off companies; iv) becoming a founder of a spin-off company. Buenstorf showed that the number of publications and citations of Max Planck scholars significantly decreased after they became founders of a spin-off company. Buenstorf's results are consistent with prior research suggesting that certain entrepreneurial behaviors, such as disclosing inventions, are complementary to academic publication and citations, while a trade-off emerges when faculty members found a spin-off firm.

Lowe and Gonzalez-Brambila (2007) are the only authors who have empirically examined the relationship between research performance and spin-off generation in both directions. In particular they analyzed whether spin-off founders are more productive compared to their colleagues, and also if founding a firm impacts on future research performance of its founding members. They concluded, analyzing a sample of 150 faculty entrepreneurs at fourteen U.S. research universities and one national laboratory who founded spin-offs between 1990 and 1999, that: i) research performance has a positive relationship with spin-off generation; and ii) at the same time there are no statistical evidences of the negative impact that founding a spin-off has on scientific performance of its academic founders.

The two studies present divergent conclusions. Despite the originality of Buenstorf's and Lowe and Gonzalez-Brambila's works, the authors themselves acknowledge that they have certain limitations. Both works analyze limited samples, therefore findings embed all the limits of inferential analysis. Buenstorf's findings are difficult to generalize since the author analyzed the scientific performance of research entrepreneurs belonging to one of the most prestigious research institutions in Europe (the Max Planck Institute): the results may considerably change when considering different universities. Also Lowe and Gonzales-Brambila's work presents limits regarding the sampling process: the study is based on a sample of 150 entrepreneurs across 15 U.S. research institutions selected according to a subjective criterion. The sample does not appear representative of the larger population of universities, since the analysis only considers few outstanding research institutions. Moreover, as demonstrated by Wong and Singh (2010) analyzing the patenting activity of leading world-wide universities, results obtained surveying US universities cannot be generalized to European universities.

However, the most severe flaw in both studies is the lack of accuracy and robustness of the methodologies employed to measure research performance, because of lack of field-standardization. Comparing research performance within such ample disciplines as chemistry, physics, biomedicine, Buenstorf himself warns the reader (page 286): "As these cross-sectional comparisons do not control for differences in the publication and citation cultures of scientific fields and disciplines, they have to be treated with caution".

Furthermore, Lowe and Gonzales-Brambila's work presents additional limits regarding the variables adopted for measurement of research performance: as indicator of productivity, the authors use the absolute number of publications, not relative with respect to their colleagues in the same discipline. The indicator of quality is whether or not the academic entrepreneur is included in the list of "high impact researchers" published by the Institute of Scientific Information (ISI). The use of this latter variable permits only the evaluation of whether the researchers that achieve spin-offs are top scientists or not. In our work we will also evaluate whether the spin-off founders' performance is greater (and how much greater) than that of their colleagues, as well as if it varies subsequent to the foundation of the spin-off, and if differences are ascertainable between the disciplines. In order to answer these questions, we carry out field-standardization of research impact for each single scientist and analyze the whole population of academic spin-offs over an eight-year period and the research performance of the whole population of research staff of the entire Italian university system. The next section provides a description.

**3. The Italian context**

Differently from other industrialized countries, Italy is characterized by research expenditures by government equal to those by the private sector. In this context of low investment by the productive sector, the exploitation of public research results by industry therefore becomes crucial for the support of the country's competitiveness. There are two structural problems though, that make the process of public-private technology transfer more difficult than in other industrialized nations. The first is the progressive hi-tech de-specialization experienced in the Italian industry in the last few years (Gallino 2003), and the second is the composition of the Italian industrial system, characterized by a disproportionate ratio of small and micro companies. The productive system has witnessed a progressive technological de-specialization, losing competitiveness in general and in the hi-tech sectors in particular (as confirmed by the performance of the indices of productive specialization, the commercial balance of payments and the technology balance), with two important effects (and/or causes). The first is the almost total disappearance of large hi-tech companies, the privileged interlocutors with the public research system, and the second is the progressive decline in private research spending. In a nutshell, with respect to other countries, in Italy, public research supply has had to face up to public support on a smaller scale together with a less "well-off" and less sophisticated private demand. The result is that a part of the results of public research cannot be absorbed into the national productive system because of mismatch (Abramo and D'Angelo 2005) and the other part finds difficulty in flowing: according to a survey by Istat (2003), the Italian companies that introduce innovations in products or processes relegate public research institutions and universities to the last positions amongst the 10 possible sources of information constituting the basis for innovation. In this context, university spin-offs represent an effective means to capture the economic returns on research expenditures and to revitalize an industrial system affected by high-tech anemia.

In 2008, a total of 87 universities were recognized by The Italian Ministry of Education, Universities and Research (MIUR), as having the authority to issue legally-recognized degrees. With only rare exceptions these are public universities largely financed through non-competitive allocation. Up to 2009, the core government funding was input oriented, i.e. distributed to universities in a manner intended to equally satisfy the needs and resources of each and all, in function of their size and activities. Further financing from the MIUR for research projects on a competitive basis represents only 9% of total income. Income deriving from technological transfer is negligible, given the very limited practice of Italian universities to carry out patenting and licensing (Abramo and Pugini 2011). All new personnel enter the university system through public examinations, and career advancement also requires such public examinations. Salaries are regulated at the nationally centralized level and are calculated according to role (administrative, technical, or professorial), rank within role (for example: assistant, associate or full professor), and seniority. No part of the salary for professors is related to merit: wages are increased annually according to parameters set by government. All professors are contractually obligated to carry out research, thus all universities are research universities: "teaching-only" universities do not exist.

Until 1996, Italian universities were characterized by: i) highly centralized governance at the national government level, and ii) low levels of autonomy. On 9 February 1996, a ministerial decree enacted Law 168/1989, which induced changes in university governance and granted universities ample margins of autonomy at strategic, financial, operational and organizational levels. This greater freedom, in particular in raising and spending financial resources, focused academic attention on licensing activities, which until then had remained negligible. It was another three years before the spin-off phenomenon was regulated by specific legislation (Decree 297, 27

July 1999)[2] . This decree came about in the context of general government policy in favor of developing innovative enterprises, and specified the characteristics under which a new venture can be defined as a research spin-off, as well as establishing ministerial funding which researchers could draw on to assist in realizing these spin-offs. According Law 297/1999, a new enterprise is defined as an academic spin-off only if it is:

- founded by university personnel with the aim of commercial benefit from academic research results;
- based on a core technology that is transferred from the parent organization;
- authorized by the originating university, which can also enter in the ownership of the spin-off company.

However, it was only with the arrival of Law 383 of 13 October 2001 (known as "Tremonti bis") that the Italian government provided true stimulus for the birth of academic spin-offs. This law establishes the right of researchers to acquire absolute title to the inventions produced by their activity in institutional role at the university. In this manner, individual inventors can freely decide whether to exploit the inventions realized, including for the establishment of their own enterprise.

Importantly, academics who establish an enterprise and personally participate in the share capital can only receive authorization to maintain their full-time professorial position (full, associate or assistant) provided that their enterprise qualifies as a university spin-off under the terms of Law 297/1999. This authorization offers numerous advantages, such as rights to use university equipment or the university logo, on the basis of royalties to the university or participation of the university in the actual capital of the spin-off.

These legal conditions are significant, if we consider that Italian researchers who found spin-offs are recognized as likely to actively participate in the enterprise management (Chiesa and Piccaluga, 2000). Further, the regulatory system for Italian academics (at least until 2010) provided that professors would have life tenure, and that their salary was calculated primarily on the basis of seniority, rather than on scientific productivity. For this reason, analysis of the relation between formation of spin-offs and scientific productivity is particularly important in the Italian context: researchers who found a spin-off, precisely because their professorial salary is not linked to results from research, could reasonably be inclined to neglect further work of that character and assign all their personal energies to their entrepreneurial activity.

---

[2] Decreto Legislativo 27 luglio 1999", http://www.camera.it/parlam/leggi/deleghe/testi/99297dl.htm. Last accessed November 28, 2011.

## 4. Dataset and methodology

### 4.1. Dataset description

The research questions refer to the population of Italian university researchers in science and engineering who founded academic spin-offs between 2001 and 2008. The universe of spin-offs was defined using the criteria of Italian Law 297 (1999),

The dataset was constructed through a survey conducted by means of interviews with administrators of the technology licensing offices of 67 Italian universities. These universities were selected on the sole criteria of having, over the period 2001-2008, at least 10 research staff employed in science and engineering. Of the 67 universities interviewed, 47 were found to have launched at least one spin-off company in the period considered. All of them are general in terms of field of research, with the sole exception of the three polytechnics (Milan, Turin and Bari), which specialize in engineering and architecture only. It should be noted that the Italian university system provides that every researcher belongs to a specific "Scientific Disciplinary Sector", or SDS. Each SDS is part of a "University Disciplinary Area", or UDA. Science and engineering are gathered in 9 UDAs[3] and 205 SDSs. For the objectives of the current work, limitation to science and engineering is necessary in order to adopt a bibliometric approach to evaluation of research performance, as described in the next section of this paper. This choice does not limit the field of investigation in a significant manner, since researcher-entrepreneurs that belong to disciplines other than science and engineering have been found to account for only 1.6% of the total.

The survey identified 326 university spin-offs[4] founded in Italy in the period under observation, from which were then excluded: i) those founded by scientists not holding a formal university faculty position[5]; ii) those where the founding members all belonged to SDSs that are not included in science and engineering.

The final dataset is composed of 284 spin-offs, originating from 47 universities based in every part of the nation, involving decidedly heterogeneous research staffs, as listed in Table 1. This large field of observation contributes to robustness of the findings as compared to previous contributions, which focus on a limited number of institutions selected from a list of the top

---

[3] Mathematics and computer sciences; physics; chemistry; earth sciences; biology; medicine; agricultural and veterinary sciences; civil engineering and architecture; industrial and information engineering.
[4] The number of start-ups generated by universities is undoubtedly higher but not all meet the criteria necessary to be defined as spin-offs.
[5] By formal faculty position we mean Italian university personnel holding a position as assistant, associate or full professor and indexed in the database of the Ministry of Universities and Research (http://cercauniversita.cineca.it/php5/docenti/cerca.php). Last accessed November 28, 2011.

institutions at the national level.

Analysis of the data by year of foundation (Table 2) shows a rapid increase in number of spin-offs from 2002 to 2004, with a subsequent stabilization, indicating an obviously delayed and prolonged effect from Law 297 of 1999, which began the regulation of the spin-off phenomenon for Italian universities, and from Law 383/2001.

The data in Table 1 show strong representation of the various polytechnics in the top positions for ranking by number of spin-offs generated. This seems consistent with the data of Table 3, which presents the distribution of spin-off founders by the UDA to which they belong. The UDA with the most observations is Industrial and information engineering (208), followed by Biology (43) and Chemistry (43). These data are not surprising, considering the highly applied nature of research activity in the engineering disciplines.

The 284 spin-offs indexed involved 427 scientists[6]. Analyzing the corporate structure of the spin-offs under examination shows that there are few cases of multiple spin-offs: only 18 scientists achieved more than one spin-off and, among these, only three achieved more than two spin-offs.

---

[6] Note that this includes all and only those researchers belonging to the nine UDAs included in the field of observation.

| University | N. of spin-offs | Research staff in science and engineering |
|---|---|---|
| Turin Polytechnic | 18 | 796 |
| University of Padua | 18 | 1,511 |
| University of Milan | 17 | 1,633 |
| Polytechnic University of the Marche | 16 | 403 |
| University of Ferrara | 14 | 512 |
| Milan Polytechnic | 13 | 1,138 |
| University of Cagliari | 12 | 755 |
| Bari Polytechnic | 11 | 334 |
| University of Modena and Reggio Emilia | 11 | 584 |
| University of Perugia | 11 | 790 |
| University of Aquila | 10 | 493 |
| University of Udine | 10 | 415 |
| University of Bologna | 10 | 1,84 |
| University of Calabria | 8 | 380 |
| University of Pisa | 8 | 1,269 |
| University of Sannio | 7 | 90 |
| University of Milan "Bicocca" | 6 | 358 |
| University of Siena | 6 | 549 |
| University of Eastern Piedmont "Amedeo Avogadro" | 5 | 170 |
| University of Salento | 5 | 256 |
| University of Florence | 5 | 1,479 |
| University of Palermo | 5 | 1,307 |
| University of Parma | 5 | 785 |
| University of Pavia | 5 | 765 |
| University of Rome "La Sapienza" | 5 | 3,126 |
| University of Rome "Tor Vergata" | 5 | 911 |
| University of Trieste | 5 | 565 |
| University of Bari | 4 | 1,08 |
| University of Messina | 4 | 925 |
| University of Camerino | 4 | 248 |
| Scuola Superiore St.Anna | 3 | 32 |
| University of Turin | 3 | 1,176 |
| Catholic University of the Holy Cross | 2 | 823 |
| University of Tuscia | 2 | 169 |
| University of Brescia | 2 | 346 |
| University of Trent | 2 | 209 |
| University of Urbino "Carlo Bo" | 2 | 152 |
| University of Verona | 2 | 344 |
| University of Rome "Tre" | 2 | 301 |
| University of Salerno | 2 | 344 |
| University of Bergamo | 1 | 55 |
| University of Foggia | 1 | 117 |
| University of Naples "Federico II" | 1 | 2,187 |
| University of Naples "Parthenope" | 1 | 68 |
| University of Magna Grecia at Catanzaro | 1 | 125 |
| University of Insubria | 1 | 230 |
| University of Catania | 1 | 1,114 |
| Total | 292* | 33,257 |

*Table 1: Distribution of spin-offs by originating university; data for 2001-2008*
*\* The total indicated differs from the total spin-offs (284) due to multiple counts concerning joint spin-offs realized by more than one university.*

| Year of foundation | N. of spin-offs | Cumulative | Cum. % |
|---|---|---|---|
| 2001 | 6 | 6 | 2,1% |
| 2002 | 8 | 14 | 4,9% |
| 2003 | 27 | 41 | 14,4% |
| 2004 | 42 | 83 | 29,2% |
| 2005 | 41 | 124 | 43,7% |
| 2006 | 47 | 171 | 60,2% |
| 2007 | 70 | 241 | 84,9% |
| 2008 | 43 | 284 | 100,0% |

*Table 2: Distribution of spin-offs in the dataset, by year of foundation*

| University Disciplinary Area | Obs |
|---|---|
| Industrial and information engineering | 208 |
| Biology | 43 |
| Chemistry | 43 |
| Civil engineering and architecture | 31 |
| Physics | 27 |
| Agricultural and veterinary sciences | 23 |
| Medicine | 22 |
| Mathematics and computer science | 21 |
| Earth sciences | 9 |
| Total | 427 |

*Table 3: Distribution of academic-entrepreneurs by scientific area*

The analysis of research performance was limited to those 382 research personnel who held formal faculty roles for at least three years over the period 2001-2008. Such researchers fall in 114 SDSs, to which belong 25,890 researchers. Table 4 presents their breakdown according to academic rank: in almost half the cases these are full professors. The data are also significant in light of the percentage incidence of full professors in the total research staff for science and engineering. The relevant concentration indexes confirm the strong concentration of full professors among spin-off founders[7]. This situation is symptomatic of the fact that researchers who have reached strong and stable positions in the university sphere are more likely to realize spin-offs.

| Academic rank (31/12/2009) | Spin-off founders | Total national research staff (only science and engineering) | Concentration index |
|---|---|---|---|
| Full professors | 185 (48.4%) | 10,543 (27.2%) | 1.78 |
| Associate professors | 116 (30.4%) | 11,401 (29.4%) | 1.03 |
| Assistant professors | 81 (21.2%) | 16,843 (43.4%) | 0.49 |
| Total | 382 | 38,787 | |

*Table 4: Distribution of academic-entrepreneurs by academic rank*

**4.2 The bibliometric approach: methodological issues**

The evaluation of scientific productivity of the researcher-entrepreneurs is based on a bibliometric approach. The literature offers ample justification for the use of scientific publications as a proxy of research output for disciplines in science and engineering (Moed et al. 2004), and of citations as a proxy of research impact (Glanzel 2008). The data used are drawn from the *Observatory on Public Research of Italy* (ORP)[8], a bibliometric database that censuses the international scientific production of all Italian public research organizations. The ORP in turn uses raw data licensed from the Thomson Reuters Web of Science ™ (WoS). Beginning from this data, and applying complex algorithms for reconciliation of the authors' affiliation and disambiguation of their true identities, each publication is attributed to the university scientists that produced it (D'Angelo et al. 2011).

---

[7] Concentration indexes represent a measure of association between two variables based on frequency data, varying around the neutral value of 1. For full professors the value of 1.78 of Table 4 derives from the ratio of two percentages in column 2 (48.4%) and 3 (27.2%).
[8] www.orp.researchvalue.it. Last accessed November 28, 2011

For each publication, an indicator of impact called Article Impact Index (AII) is considered. AII is given by the ratio between the number of citations received by the publication and the median[9] of citations of all Italian publications in the same year and the same WoS subject category. For publications in multi-category journals, AII is calculated as a weighted average of the values referring to the individual subject categories. Standardizing citations by the median, the well-known distortions typical of measurements which lack field-standardization have been avoided. The magnitude of the distortions that occur when indicators are measured at the aggregate level may be very relevant[10] (Abramo et al. 2008). To the authors' knowledge, the ORP approach is the only model of research assessment in the world that is able to provide bibliometric national-scale field-standardized evaluations at the individual level.

The evaluation of bibliometric performance by individual researchers is based on the following indicators:

- **Output**, O. Sum of the publications[11] produced by a researcher over the period considered.
- **Fractional output**, FO. Sum of the publications produced by a researcher, each one weighted according to: i) number of co-authors; and, in the case of the life sciences, ii) the position of the author in the list, iii) character of the co-authorship (intra-mural or extra-mural).
- **Quality index**, QI. Average impact of the publications of a researcher, given by the average value of their respective Article Impact Index.
- **Scientific strength**, SS. Product of the Output (O) and the Quality index (QI) of a given researcher.
- **Fractional scientific strength,** FSS. As for fractional output, but referring to Scientific Strength.

---

[9] Publications without citations are excluded from calculation of the median. We use the median rather than the world average provided by WoS because of the skewness of citation distributions (Lundberg, 2007).
[10] To provide an example, within the discipline of biology, an article falling in the field of biochemistry receives on average around 24 citations after eight years, while an article in mycology around seven. Without standardizing by field, it would be hard for a mycologist to show a higher impact than a biochemist.
[11] Only articles, article reviews and conference proceedings.

## 5. Results

In this section we present the results from elaborations intended to answer the original research questions.

*Do faculty members who found spin-off ventures have higher research performance than their colleagues (i.e. publish more and higher impact papers?)*

To respond to this first question, absolute values of all the performance indicators were calculated for each of the 382 observed researchers. These values were then averaged by the number of years that each researcher actually held an official faculty role. For comparison with the same data referring to all national academic researchers in the same SDS (not just a control group), the performance was expressed as a national percentile (0 lowest value, 100 best).

This procedure takes account of the variability in intensity of publication and citation in different fields of science and engineering and thus, by avoiding field distortion, permits the robust comparison of the bibliometric performance of scientists working in very different SDSs (different fields of research). Moreover, percentile values embed comparison among scientists: a value above 50 means performance above average and vice versa.

A first evaluation, obtained grouping the researchers by disciplinary area, shows that in all disciplinary areas, researcher-entrepreneurs demonstrate better performance than their remaining national colleagues in the same field (Table 5)[12]. In terms of output (O), the average percentile of performance by researcher-entrepreneurs is always significantly higher than that by non-entrepreneurs, varying from a minimum of 55.5 for Civil engineering and architecture to a maximum of 81.0 for Biology. Significance is borderline only for Earth Sciences. Similar superiority is seen for all other performance indicators, with significance not shown only in Mathematics and, in the case of the average quality (QI) indicator, in Earth Sciences and Industrial and information engineering.

We also investigated if differences in performance vary across academic rank: we classified researcher-entrepreneurs into three academic ranks (full, associate and assistant professors), and compared their performance with their complements in the same rank. Findings are shown in

---

[12] The low values of average percentile for performance indicators of non-entrepreneurial researchers in Civil engineering and architecture are due to the very large proportion of non-productive researchers (0 percentile) in that discipline.

Table 6.

| University Disciplinary Area | | Obs | O | FO | QI | SS | FSS |
|---|---|---|---|---|---|---|---|
| Mathematics and computer science | Entrepreneurs | 20 | 62.1 | 58.4 | 50.7 | 56.4 | 56.0 |
| | Others | 1,901 | 48.5 | 47.4 | 44.8 | 44.8 | 44.8 |
| | *Test t (p-value)* | | *0.030* | *0.062* | *0.220* | *0.065* | *0.071* |
| Physics | Entrepreneurs | 27 | 74.1 | 73.4 | 64.4 | 72.8 | 71.0 |
| | Others | 2,010 | 50.1 | 49.4 | 49.0 | 49.0 | 49.0 |
| | *Test t (p-value)* | | *0.000* | *0.000* | *0.000* | *0.000* | *0.000* |
| Chemistry | Entrepreneurs | 41 | 76.1 | 74.8 | 57.3 | 71.6 | 70.4 |
| | Others | 3,362 | 50.6 | 49.9 | 49.9 | 49.9 | 49.9 |
| | *Test t (p-value)* | | *0.000* | *0.000* | *0.052* | *0.000* | *0.000* |
| Earth sciences | Entrepreneurs | 9 | 67.1 | 71.4 | 52.4 | 57.5 | 60.2 |
| | Others | 668 | 50.5 | 49.0 | 47.8 | 47.8 | 47.8 |
| | *Test t (p-value)* | | *0.053* | *0.015* | *0.335* | *0.185* | *0.125* |
| Biology | Entrepreneurs | 37 | 81.0 | 78.6 | 70.4 | 80.6 | 78.8 |
| | Others | 4,119 | 50.6 | 49.6 | 49.4 | 49.4 | 49.4 |
| | *Test t (p-value)* | | *0.000* | *0.000* | *0.000* | *0.000* | *0.000* |
| Medicine | Entrepreneurs | 19 | 78.0 | 75.5 | 67.8 | 77.2 | 75.7 |
| | Others | 4,431 | 49.9 | 49.3 | 48.9 | 48.9 | 48.9 |
| | *Test t (p-value)* | | *0.000* | *0.000* | *0.003* | *0.000* | *0.000* |
| Agricultural and veterinary sciences | Entrepreneurs | 21 | 71.3 | 71.3 | 57.6 | 65.1 | 65.7 |
| | Others | 2,063 | 46.4 | 45.1 | 41.7 | 41.7 | 41.7 |
| | *Test t (p-value)* | | *0.000* | *0.000* | *0.020* | *0.001* | *0.001* |
| Civil engineering and architecture | Entrepreneurs | 28 | 55.5 | 53.9 | 47.2 | 47.3 | 47.6 |
| | Others | 2,403 | 33.9 | 33.1 | 25.9 | 25.9 | 25.9 |
| | *Test t (p-value)* | | *0.001* | *0.002* | *0.001* | *0.001* | *0.000* |
| Industrial and information engineering | Entrepreneurs | 180 | 61.6 | 59.8 | 49.1 | 53.8 | 53.2 |
| | Others | 4,933 | 49.9 | 49.0 | 46.6 | 46.6 | 46.6 |
| | *Test t (p-value)* | | *0.000* | *0.000* | *0.161* | *0.002* | *0.004* |
| Total/Weighted Average | | 382 | 67.0 | 65.4 | 54.5 | 61.2 | 60.5 |

***Table 5: Distribution by UDA for the average percentile of performance by researcher-entrepreneurs and non-entrepreneurs, with significance levels of differences by t test.***

| Indicator | | Full professors | Associate professors | Assistant professors |
|---|---|---|---|---|
| O | Entrepreneurs | 73,0 | 64,3 | 57,2 |
| | Others | 55,5 | 45,1 | 43,4 |
| | t test | -7.482*** | -6.543*** | -4.172*** |
| FO | Entrepreneurs | 71,9 | 63,4 | 53,4 |
| | Others | 55,1 | 44,8 | 41,4 |
| | t test | -7.132*** | -6.341*** | -3.688*** |
| QI | Entrepreneurs | 60,5 | 52,6 | 43,7 |
| | Others | 50,8 | 44,1 | 41,8 |
| | t test | -4.115*** | -2.802** | -0.533 |
| SS | Entrepreneurs | 68,9 | 58,7 | 47,2 |
| | Others | 53,1 | 43,0 | 40,1 |
| | T test | -6.460*** | -5.187*** | -2.065* |
| FSS | Entrepreneurs | 68,3 | 59,1 | 44,8 |
| | Others | 53,6 | 43,4 | 39,2 |
| | T test | -6.012*** | -5.156*** | -1.665* |

***Table 6: Comparison of performance between researcher-entrepreneurs and non-entrepreneurs per academic rank***
*\*\*\*: p value <0.001*
*\*\*: p-value <0.01*
*\*: p-value < 0.05*

The next step in the analysis was to calculate the index of concentration of researcher-entrepreneurs among the top national scientists[13]. This index allows determination of whether, and in what measure, the researcher-entrepreneurs (which the preceding results have just shown to be, on average, more productive than their national colleagues) place in the group of most distinguished scientists.

The findings are presented in Table 7. In general we see that 27% of researcher-entrepreneurs place among top Italian scientists for number of publications (O) and 25% of them place among the top for impact (FSS). Biology is the area with the greatest concentration of researcher-entrepreneurs among its top scientists: roughly 50% for each dimension of performance, with the exception of QI (27%). Concentrations of entrepreneurs among top scientists are also particularly high in Physics and in Medicine (37% for output). On the opposite side, in Earth Sciences there was no researcher-entrepreneur with top performance by output (O) or scientific strength (SS) concentration.

| University Disciplinary Area | Obs | Output | Fractional Output | Quality Index | Scientific Strength | Fractional Scientific Strength |
|---|---|---|---|---|---|---|
| Mathematics and computer science | 20 | 2.5** | 2.5** | 0.5 | 2.0 | 2.0 |
| Physics | 27 | 3.7*** | 2.6** | 1.5 | 3.3*** | 3.0*** |
| Chemistry | 41 | 3.4*** | 3.7*** | 1.7 | 2.9*** | 3.2*** |
| Earth sciences | 9 | 0.0 | 1.1 | 2.2 | 0.0 | 2.2 |
| Biology | 37 | 4.9*** | 4.9*** | 2.7*** | 5.4*** | 4.9*** |
| Medicine | 19 | 3.7*** | 3.7*** | 1.1 | 3.7*** | 3.7*** |
| Agricultural and veterinary sciences | 21 | 3.3*** | 3.8*** | 1.0 | 1.9 | 2.9** |
| Civil engineering and architecture | 28 | 2.1** | 1.4 | 2.5** | 1.8 | 1.8 |
| Industrial and information engineering | 180 | 2.1*** | 1.8*** | 0.8 | 1.7** | 1.8*** |
| Total/Weighted Average | 382 | 2.7 | 2.5 | 1.3 | 2.4 | 2.5 |

*Table 7: Distribution by UDA for indexes of concentration of researcher-entrepreneurs among top national scientists and significance levels of differences from 1.*
*\*\*\*: p value <0.001*
*\*\*: p-value <0.01*
*\*: p-value < 0.05*

Finally, referring to numerosity of spin-offs achieved, it is useful to compare the performance of those who contribute to founding a single spin-off with those who initiate more than one. From the data in Table 8 it would seem that the average performance of the members of these two groups does not differ in a significant manner. However, analyzing the data for those 18 researchers who did contribute to more than one spin-off, it is interesting to note that: one third (6) fall in the group of top scientists (national percentile above 90%), but of the 12 remaining, eight had scientific performance under the national median. Further, the two scientists with three spin-offs had a thoroughly mediocre performance.

---

[13] "Top scientist" is defined as a researcher with a performance among the top 10% in the nation for his/her SDS (a percentile equal to or greater than 90). An index of concentration of 1.4 indicates that, out of the total of all researcher-entrepreneurs, 14% are "top scientists".

In summary, referring to the first research question, the findings show better scientific performance by researcher-entrepreneurs when compared to their national colleagues in the same SDS.

| Indicator / Obs | Only one 364 | Two or more 18 | p-value[*] |
|---|---|---|---|
| Output | 67.1 | 65.0 | 0.982 (=) |
| Fractional Output | 65.5 | 62.4 | 0.866 (=) |
| Quality Index | 54.4 | 57.8 | 0.905 (=) |
| Scientific Strength | 61.2 | 60.6 | 0.989 (=) |
| Fractional Scientific Strength | 60.6 | 59.1 | 0.452 (=) |

*Table 8: Average percentile of performance for researcher-entrepreneurs; data by number of spin-offs realized*
[*] NPC test

The second research question is:

*Does research performance of entrepreneurial academics, contrasted to that of their colleagues, decrease after the founding of spin-off ventures?*

To respond to this question an inter-temporal analysis has been applied. Calculations were made for each individual researcher, using the same indicators as above, of scientific performance in the triennium preceding and the triennium subsequent to the creation of a spin-off. The analysis deals only with spin-offs achieved between 2004 and 2006 because these are the only years for which it is possible to conduct the two triennial analyses. Thus the number of observations is reduced to 131, subdivided as indicated in Table 9.

| Year of spin-off foundation | Time series considered | Obs |
|---|---|---|
| 2004 | 2004-2006 vs. 2001-2003 | 33 |
| 2005 | 2005-2007 vs. 2002-2004 | 53 |
| 2006 | 2006-2008 vs. 2003-2005 | 45 |
| | Total | 131 |

*Table 9: Observations included in time-series performance analysis*

Calculations were then made for the difference in performance percentile between the triennium following and that preceding the foundation of the spin-off. Table 10 shows the results for average differences and relevant statistics, per UDA. Overall, there are minor variations at the level of output (+0.09) and a slight worsening in terms of average quality (-2.73) and total impact (FSS, -2.25). However there is a certain heterogeneity among the various areas. In Industrial and information engineering the difference between the two triennia is negative for all the indicators, falling between -2.22 percentiles for SS and -3.72 percentiles for FO. In Chemistry the variation in performance over time also results as negative for all indicators. On the opposite side, for the 10 observations in Biology, the average difference in performance between the two triennia is always positive, falling between a minimum of +4.21 for FSS and a maximum of +8.17 for SS. The other areas do not show significant situations, probably in part due to the limited number of observations available. In some areas there are diverging results: output increases while impact decreases, or vice versa. Disregarding distinction by ADU, the distribution of relative differences for all five indicators appears significantly symmetrical (skewness between -0.26 and -0.84) and with non-nil median only seen for Scientific Strength (-0.48). Both the paired t test (last line in Table 10) and the Wilcoxon test, which is less sensitive to outliers, confirm the null hypothesis, meaning that there is no variation in performance between before and after the founding of a spin-off. Table 11 presents a detailed analysis showing the number of observations that are positive (+), negative (-) or nil (=), for the difference in performance by researcher-entrepreneurs between the triennium following and that prior to foundation of a spin-off. The table shows, at the overall level, that output does not worsen in 56% of cases (74 of 131), while average quality worsens in 52% of cases (68 of 131). The researcher-entrepreneurs that increase collaboration are more in number than those that decrease collaboration; in fact there are more negative shifts in FO (63) than there are positive (54) after foundation of spin-offs. At the level of the individual area, the data show more improvements than declines in output for Civil engineering and architecture. The average quality (QI) shows numerous cases of perceptible improvement in Physics and in Agricultural and veterinary sciences, but equally numerous instances of worsening in Mathematics and computer science, Chemistry, and Civil engineering and architecture. In Industrial and information engineering and Biology there are more cases where overall impact (FSS) improves than where it worsens, while the contrary takes place in Mathematics and computer science, Chemistry and Medicine.

| UDA | Obs | Output | Fractional Output | Quality Index | Scientific Strength | Fractional Scientific Strength |
|---|---|---|---|---|---|---|
| Mathematics and computer science | 13 | +3.19 | +4.96 | -8.55 | -5.69 | -1,71 |
| Physics | 12 | -0.27 | -1.56 | +8.46 | +3.99 | +2,02 |
| Chemistry | 18 | -0.62 | -1.03 | -13.15 | -4.00 | -6,24 |
| Earth sciences | 2 | +4.16 | +4.22 | -7.59 | -3.10 | -3,55 |
| Biology | 10 | +7.11 | +4.89 | +6.58 | +8.17 | +4,21 |
| Medicine | 7 | +0.13 | -5.00 | +2.79 | +1.65 | -2,88 |
| Agricultural and veterinary sciences | 8 | -6.02 | -10.92 | +17.69 | +5.79 | +2,58 |
| Civil engineering and architecture | 13 | +6.88 | 4.87 | -11.20 | -9.64 | -8,47 |
| Industrial and information engineering | 48 | -2.87 | -3.72 | -3.70 | -2.22 | -2,30 |
| Total/Weighted average | 131 | +0.09 | -1.17 | -2.73 | -1.50 | -2,25 |
| Median | | 0 | 0 | 0 | -0.48 | 0 |
| St. Dev. | | 19,1 | 21.02 | 25.32 | 30.19 | 24.63 |
| Max | | 64,53 | 58.14 | 79.79 | 94.68 | 76.6 |
| Min | | -71,8 | -76.5 | -96.4 | -99.1 | -96.9 |
| Kurtosis | | 3,476 | 1.997 | 3.95 | 2.504 | 4.214 |
| Skewness | | -0,27 | -0.26 | -0.81 | -0.35 | -0.84 |
| Paired t test | | -0.052 | 0.636 | 0.679 | 1.047 | 1.036 |
| (p value) | | (0.959) | (0.526) | (0.498) | (0.297) | (0.302) |

*Table 10: Distribution by UDA and relevant statistics of average difference in percentile of performance by researcher-entrepreneurs, between the triennia subsequent to and preceding foundation of a spin-off*

| | Output | | | Fractional Output | | | Quality Index | | | Scientific Strength | | | Fractional Scientific Strength | | |
|---|---|---|---|---|---|---|---|---|---|---|---|---|---|---|---|
| UDA | + | = | - | + | = | - | + | = | - | + | = | - | + | = | - |
| Mathematics and computer science | 5 | 2 | 6 | 7 | 2 | 4 | 2 | 2 | 9 | 4 | 2 | 7 | 4 | 2 | 7 |
| Physics | 6 | 1 | 5 | 5 | 1 | 6 | 8 | 1 | 3 | 8 | 1 | 3 | 6 | 1 | 5 |
| Chemistry | 8 | 0 | 10 | 9 | 0 | 9 | 4 | 0 | 14 | 7 | 0 | 11 | 5 | 0 | 13 |
| Earth sciences | 1 | 0 | 1 | 1 | 0 | 1 | 0 | 0 | 2 | 1 | 0 | 1 | 1 | 0 | 1 |
| Biology | 6 | 0 | 4 | 6 | 0 | 4 | 4 | 1 | 5 | 5 | 1 | 4 | 6 | 1 | 3 |
| Medicine | 4 | 0 | 3 | 1 | 0 | 6 | 3 | 0 | 4 | 3 | 0 | 4 | 2 | 0 | 5 |
| Agricultural and veterinary sciences | 3 | 0 | 5 | 3 | 0 | 5 | 8 | 0 | 0 | 6 | 0 | 2 | 3 | 0 | 5 |
| Civil engineering and architecture | 6 | 5 | 2 | 4 | 5 | 4 | 1 | 6 | 6 | 2 | 6 | 5 | 3 | 6 | 4 |
| Industrial and information engineering | 21 | 6 | 21 | 18 | 6 | 24 | 17 | 6 | 25 | 21 | 8 | 19 | 23 | 8 | 17 |
| Total | 60 | 14 | 57 | 54 | 14 | 63 | 47 | 16 | 68 | 57 | 18 | 56 | 53 | 18 | 60 |

*Table 11: Distribution by UDA of number of observations for difference in performance that are positive (+), negative (-) or nil (=) by researcher-entrepreneurs, between the triennia subsequent to and preceding foundation of a spin-off.*

Findings show that after development of a spin-off company the number of academic founders who worsen their performance is about the same as those who improve it, with negligible differences among disciplines. In terms of number of publications (O) and overall impact (SS), the average performance slightly improves. The opposite occurs for the fractional indicators, meaning that after founding a spin-off the researcher-entrepreneur tends to collaborate more in his/her research activities. Because changes in the working environment, if any, have occurred for both founders and non-founders, we can conclude that, contrary to what emerges in Buenstorf's study (2009), involvement in the creation of a spin-off does not seem to have an overall negative effect on the scientific performance of the academic founders.

In answering both research questions, inferential analysis was not applied, because the dataset is made up of all the spin-off founder population, within the academic population in science and engineering. Furthermore, the authors warn about generalizing these findings to other national contexts, because academic entrepreneurship is strongly affected by organizational culture, national laws, policies and management systems. However the authors acknowledge that the borderline between populations and samples is sometimes blurred; for this reason they supplied some tables with t test values.

## 5. Discussion

The phenomenon of creating new enterprises by processes of spinning off from universities has attracted increasing attention over the years, from both policy-makers and scholars. This is because of the positive impact that such spin-off enterprises can have on economic development in the local area. One of the controversial aspects of university staff participating in such spin-offs concerns the relationship between the entrepreneurial activities and the research activities of the individual researcher-entrepreneurs.

The current work is an attempt to deepen the few existing studies and assist in clarifying the relationship that exists between spin-offs and scientific performance, both verifying if researcher-entrepreneurs have research productivity that is better than that of their non-entrepreneur colleagues, and evaluating if and to what extent the involvement in entrepreneurial activity influences research productivity of the individuals.

The analyses concerning the first research question showed that, in all the disciplines considered, academic entrepreneurs show research performance that is on average better than that of their national colleagues. This holds true regardless of the research field considered or the academic rank of the researcher-entrepreneurs. The results also confirm and support existing studies (Lowe and Gonzales-Brambila, 2007) concerning the hypothesis that researchers who found spin-offs are significantly concentrated among star scientists. The methodology applied also permitted measurement of the extent to which researcher-entrepreneurs perform better than their other national colleagues. This spread is greatest in Biology, a fact that confirms previous studies suggesting that university-to-firm technology transfers of breakthrough invention in biotechnology typically involve star scientists (Zucker and Darby 2007; Zucker et al. 2002; Zucker et al. 1998). Finally, analysis of indicators of productivity for entrepreneurs who realize more than one spin-off shows similar results to those reported by Fabrizio e Di Minin (2008), for patents, and by Bonaccorsi et al. (2006), for private funds: entrepreneurial and academic activity are not in conflict, and it is only in certain circumstances, when entrepreneurship is exaggerated, that scientific performance suffers.

The results concerning the second research question confirms even more clearly that the creation of spin-offs does not impact negatively on scientific performance of academic entrepreneurs. Such entrepreneurs not only have scientific performance that is higher than that of their colleagues, but this spread also does not modify even after the creation of the spin-off, and in some cases the scientific performance of the entrepreneurial researcher actually improves. These results are in contrast to those claimed by Buenstorf (2009), while confirm the conclusions of Lowe and Gonzales-Brambila's work (2007). However, compared to this latter study, our own is based on the entire population of Italian academic entrepreneurs. Finally, unlike previous studies, the current analyses include both high and low reputation universities and thus offer a greater level of generalizability compared to the contribution by Lowe and Gonzales-Brambila (2007). In an improvement over Lowe and Gonzales-Brambila, the scientific performance of researcher-entrepreneurs was evaluated using a bibliometric approach that is particularly sophisticated and robust, as described by Abramo and D'Angelo (2011). Applying this approach, the comparison of performance by individuals with respect to national distributions in the same field in a robust way permits not only verification of whether the researcher-entrepreneurs are top scientists, but also of how much their productivity is greater than that of others.

Last but not least, the specificity of the Italian situation provides convincing support for the numeric results and conclusions and guarantees their generalizability. In Italy, faculty members have life tenure and their remuneration is primarily linked to seniority, and not to scientific productivity. For this reason, even more so than in other contexts, researcher-entrepreneurs could reasonably be led to neglect their institutional responsibilities, particularly their research, and devote all their energies to their enterprise. If this does not occur in Italy then the argument is still stronger that it should certainly not occur in nations where there is an incentive system linked to scientific productivity.

5. Conclusions

In summary, the thesis that emerges from this work is that entrepreneurship and scientific research by academic scholars are not in conflict, even when entrepreneurialism takes its most extreme form in the realization of spin-offs.

The work undertakes to contribute to studies on the spin-off phenomenon and the concept of development of the entrepreneurial university, analyzing an aspect that has been little examined in the literature: the relation between creation of spin-offs and the scientific performance of the academic founders. The paper, contrary to what has been feared by some scholars, demonstrates that both creation of enterprise and scientific activity can tranquilly coexist in the agendas of individual scientists. The founders of spin-offs are on average more productive than their colleagues in the same field of research, and the creation of their spin-offs does not determine a reduction in their scientific productivity. The current study presents a number of original elements compared to previous research. To the best of our knowledge, it is the only study so far able to empirically demonstrate with high level of statistical significance that spin-off creation does not cause a decrease in scientific productivity of its academic founders. The research questions were verified by examining the cohort of all Italian university researchers that participated in creating a university spin-offs between 2001 and 2008. The focus on Italy permitted analysis of the spin-off/scientific productivity relationship in a nation where university entrepreneurial activity has only recently been regulated by specific policy interventions, intended to provide incentives. This latter point constitutes an innovative aspect compared to the studies present in the literature, which refer to university systems in nations that clearly have a more significant university entrepreneurial tradition (particularly USA, Canada, and Germany). Moreover, the specificity of the Italian context provides convincing support for the results and conclusions and should guarantee their generalization to other nations. In Italy, faculty members have life tenure from the very beginning; and their remuneration is primarily linked to seniority, and not to productivity. For this reason, even more so than in other contexts, entrepreneurial researchers could reasonably neglect their institutional responsibilities, particularly their research mission, and devote all energies to their enterprise. If this does not occur in Italy, then it should be even more so in other nations where career tracks and incentive systems are strongly linked to research performance.

Further, a large part of the existing studies on the issue tend to concentrate their analysis either on a single disciplinary sector, or on a number of disciplines but at few research institutions. The dataset used in this paper, in contrast, refers to the entire Italian university system, without sectorial or institutional limitations (except for choices concerning significance). Finally, and most importantly, the productivity measurements conducted in this study are field-standardized, avoiding the distortions affecting previous studies on the subject.The scientific performance of researcher-entrepreneurs was evaluated using a bibliometric approach that is particularly sophisticated and robust, as described by Abramo and D'Angelo (2011). Applying this approach, the comparison of performance by individuals with respect to national distributions in the same field is free from distortions due to lack of field-standardization and permits not only verification of whether the researcher-entrepreneurs are top scientists, but also of how much their productivity is greater than that of others.

Finally our paper gives important indications for university administrators and policy makers.

If the birth of new technology-based firms through spin-off from universities seems desirable for the positive impact that these firms can have on economic development in the area, it is also true that many policy makers and university administrators may fear the impact of entrepreneurial activities on university research productivity.

Obstacles to development of university entrepreneurship often actually arise within the universities themselves, posed by those who hold that entrepreneurship and research activities are incompatible, and supported by studies that indicate a negative impact of the entrepreneurial activity on research. This aversion to academic entrepreneurship is more evident in Italy. Despite an increasing focus of policy interventions aimed at fostering academic entrepreneurship, the share of public funds allocated to Italian universities on the basis of merit, is still determined according to the teaching and the research performances. This element favors the diffusion among academics of a "publish or perish" philosophy and induce universities' deans, anxious to maximize universities funding, to neglect any forms of academic entrepreneurship.

Instead, this paper provides an opposite indication, demonstrating that the birth of spin-offs does not prejudice research activity by individual scientists. This can encourage: i) universities to promote academic entrepreneurship, because of the absence of a negative impact on scientific productivity and consequently on the competition for funds; and ii) policy maker to promote more initiatives to foster academic entrepreneurship.

While already shedding new light on the relationship between spin-off realization and research performance of academic entrepreneurs, this study also offers a number of points of departure for deeper consideration. The model for evaluation of scientific performance could integrate variables of context (organizational, geographic, etc.), or some other spin-off characteristics (inventions licensed to the spin off, equity ownership, etc.) which have not been considered here. In addition, when further historic data series are available, it would also be interesting to examine the relation of scientific productivity of individual researchers not only with the birth of academic spin-offs, but also with their long-term survivability.

**References**

Abramo, G. and D'Angelo, C.A. (2011). National-scale research performance assessment at the individual level. *Scientometrics*, 86(2), 347-364.

Abramo, G., D'Angelo, C.A. and Di Costa, F. (2009). Research collaboration and productivity: is there correlation? *Higher Education*, 57(2), 155-171.

Abramo, G., D'Angelo, C.A. and Di Costa, F. (2008). Assessment of sectoral aggregation distortion in research productivity measurements. *Research Evaluation*, 17(2), 111-121.

Abramo, G. and Pugini, F. (2011). Assessing the relative technology transfer performance of universities and public research laboratories: the case of Italy. Forthcoming in *International Journal of Technology Transfer and Commercialisation*.

Adams, J. and Griliches Z. (1998). Research productivity in a system of universities. *Annales d'economie et de statistique*, 49/50, 127-162.

Agrawal, A. and Henderson, R. (2002). Putting patents in context: exploring knowledge transfer from MIT. *Management Science*, 48(1), 44–60.

Azoulay, P., Ding, W. and Stuart, T. (2006). The impact of academic patenting on the rate, quality and direction of (public) research. *NBER working paper* 11917, Cambridge, MA.

Benneworth, P. and Charles, D. (2005). University spin-off policies and economic development in Less successful regions: Learning from two decades of policy practice. *European Planning Studies*, 13(4), 537 – 557.

Bonaccorsi, A., Daraio, C. and Simar, L. (2006). Advanced indicators of productivity of universities. An application of robust nonparametric methods to Italian data. *Scientometrics*, 66 (2), 389–410.

Breschi, S., Lissoni, F. and Montobbio, F. ( 2007). The scientific productivity of academic inventors: new evidence from Italian data. *Economics of Innovation and New Technology*, 16(2), 101–118.

Buenstorf, G. (2009). Is commercialization good or bad for science? Individual-level evidence from the Max Planck Society. *Research Policy*, 38, 281-292.

Carayol, N. (2007). Academic incentives, research organization and patenting at a large French university. *Economics of Innovation and New Technology* 16 (2), 71–99.

Chang, Y.C. and Yang, P.Y. (2008). The impacts of academic patenting and licensing on knowledge production and diffusion: a test of the anti-commons effect in Taiwan, *R&D Management*, 38 (3), 321-334.

Chang, Y.C., Yang, P.Y. and Chen, M.H. (2009). The determinants of academic research commercial performance: Towards an organizational ambidexterity perspective. *Research Policy*, 38, 936-946.

Chiesa, V. and Piccaluga, A. (2000). Exploitation and diffusion of public research: the case of academic spin-off companies in Italy. *R&D Management*, 30(4), 329-339.

Czarnitzki, D., Glänzel W. and Hussinger, K. (2007). Patent and publication activities of German professors: an empirical assessment of their co-activity. *Research Evaluation*, 16 (4), 311–319.

D'Angelo, C.A., Giuffrida, C. and Abramo, G. (2011). A heuristic approach to author name disambiguation in large-scale bibliometric databases. *Journal of the American Society for Information Science and Technology*, 62(2), 257–269.

Dasgupta, P. and David, P.A. (1987). Information disclosure and the economics of science and technology. In: Feiwel, G.R. (Ed.), *Arrow and the Ascent of Modern Economic Theory*. NY University Press, New York, pp. 519–542.

Dasgupta, P. and David, P.A. (1994). Toward a new economics of science. *Research Policy*, 23, 487–521.

Davenport, S., Carr, A. and Bibby, D. (2002). Leveraging talent: spinoff strategy at industrial research. *R&D Management*, 32 (3), 241-254.

Debackere, K. and Veugelers, R. (2005). The role of academic technology transfer organizations in improving industry-science links. *Research Policy*, 34, 321-342.

Di Gregorio, D. and Shane, S. (2003). Why do some universities generate more start-ups than others? *Research Policy*, 32, 209-227.

Etzkowitz, H. (2003). Research groups as 'quasi-firms': the invention of the entrepreneurial university, *Research Policy*, 32, 109–121.

Fabrizio, K.R. and Di Minin, A. (2008). Commercializing the laboratory: Faculty patenting and the open science environment. *Research Policy*, 37, 914-931.

Feller, I. (1990). Universities as engines of R&D-based economic growth: they think they can. *Research Policy*, 19, 335–348.

Florida, R. and Cohen, W.M. (1999). Engine or infrastructure? The university role in economic development. in: Branscomb L.M., Kodama F., Florida R. (Eds.), *Industrializing Knowledge: University–Industry Linkages in Japan and the United States*. MIT Press, London, pp. 589–610.

Gallino, L. (2003). *La scomparsa dell'Italia industriale*, Giulio Einaudi Editore. ISBN 9788806166281.

Glänzel, W. (2008). Seven myths in bibliometrics. About facts and fiction in quantitative science studies. Kretschmer & F. Havemann (Eds): *Proceedings of WIS Fourth International Conference on Webometrics, Informetrics and Scientometrics & Ninth COLLNET Meeting*, Berlin, Germany.

Gulbrandsen, M. and Smeby, J.C. (2005). Industry funding and university professors' research performance. *Research Policy,* 34, 932–950.

Hane, G. (1999). Comparing university–industry linkages in the United States and Japan. In Branscomb, L.M., Kodama, F., Florida, R. (Eds.), *Industrializing Knowledge: University–Industry Linkages in Japan and the United States*. MIT Press, London, pp. 20–61.

Istat (2003). L'innovazione nelle imprese italiane 1998-2000.

Jacobsson, S. ( 2002). Universities and industrial transformation. *Science and Public Policy*, 29 (5), 345–365.

Lach, S. and Schankerman, M. (2003). Incentives and invention in universities, *National Bureau of Economic Research*, working paper 9727.

Landry, R., Amara, N. and Rherrad, I. (2006). Why are some university researchers more likely to create spin-offs than others? Evidence from Canadian universities. *Research Policy*, 35, 1599–1615.

Larsen, M.T. (2011). The implications of academic enterprise for public science: An overview of the empirical evidence. *Research Policy*, 40, 6-19.

Louis, K.S., Blumenthal, D., Gluck, M.E. and Stoto, M.A. (1989). Entrepreneurs in Academe: An exploration of behaviours among life scientists. *Administrative Science Quarterly*, 34, 110-131.

Lowe, R.A. and Gonzalez-Brambila, C. (2007). Faculty entrepreneurs and research productivity. *Journal of Technology Transfer*, 32, 173–194.

Lundberg, J. (2007). Lifting the crown-citation z-score. *Journal of Informetrics*, 1(2), 145–154.

Manjarrés-Henríquez, L., Gutiérrez-Garcia, A. and Vega-Jurado, J. (2008). Coexistence of

university–industry relations and academic research: Barrier to or incentive for scientific productivity. *Scientometrics*, 76(3), 561–576.

Meyer, M. (2006). Academic inventiveness and entrepreneurship: on the importance of start-up companies in commercializing academic patents. *The Journal of Technology Transfer*, 31 (4), 501–510.

Metcalfe, J.S. (1998). *Evolutionary Economics and Creative Destruction*. Routledge, London and New York.

Moed, H.F., Glänzel, W. and Schmoch, U. (2004). *Handbook of Quantitative Science and Technology Research: The Use of Publication and Patent Statistics in Studies of S & T Systems*. Springer.

Murray, F. (2004). The role of academic inventors in entrepreneurial firms: Sharing the laboratory life. *Research Policy*, 3, 643–659.

Nelson, R.R. (1959). The simple economics of basic scientific research. *Journal of Political Economy*, 67 (3), 297–306.

Nelson, R.R. (2001). Observations on the post-Bayh-Dole rise of patenting at American universities. *Journal of Technology Transfer*, 26 (1–2), 13–19.

O'Shea, R.P., Allen, T.J., Morse, K.P., O'Gorman, C. and Roche, F. (2007). Delineating the anatomy of an entrepreneurial university: the Massachusetts Institute of Technology experience. *R&D Management*, 37(1), 1-16.

O' Shea, R.P., Chugh, H. and Allen, T.J. (2008). Determinants and Consequences of University Spin-off Activity: A Conceptual Framework. *Journal of Technology Transfer*, 33, 653-666.

Powers, J.B. and McDougall, P.P. (2005). University start-up formation and technology licensing with firms that go public: a resource-based view of academic entrepreneurship. *Journal of Business Venturing*, 20, 291-311.

Rothaermel, F.T. and Thursby, M. (2005). Incubator firm failure or graduation? The role of university linkages. *Research Policy*, 34, 1076 –1090.

Shane, S. (2004a). *Academic entrepreneurship: University spin-offs and wealth creation*. Cheltenham, UK: Edward Elgar.

Shane, S. (2004b). Encouraging university entrepreneurship? The effect of the Bayh-Dole act on university patenting in the United States, *Journal of Business Venturing*, 19(1), 127–151.

Stephan, P., Gurmu, S., Sumell, A.J. and Black, G. (2007).Who's patenting in the university?

*Economics of Innovation and New Technology*, 16(2), 71–99.

Toole, A.A. and Czarnitzki, D. (2010). Commercializing Science: Is there a university "brain drain" from academic entrepreneurship? *Management Science*, 56 (9), 1599-1614.

Thursby, J.G. and Kemp, S. (2002). Growth and productive efficiency of university intellectual property licensing. *Research Policy*, 31 (1), 109-124.

Van Looy, B., Callaert, J. and Debackere, K. (2006). Publication and patent behaviour of academic researchers: conflicting, reinforcing or merely co-existing?. *Research Policy*, 35(4), 596–609.

Van Looy, B., Ranga, M., Callaert, J., Debackere, K. and Zimmermann, E. (2004). Combining entrepreneurial and scientific performance in academia: toward a compounded and reciprocal Matthew-effect? *Research Policy*, 33, 425-441.

Van Looy, B., Landoni, P., Callaert, J., van Pottelsberghe, B., Sapsalis, E. and Debackere, K. (2011). Entrepreneurial effectiveness of European universities: An empirical assessment of antecedents and trade-offs, *Research Policy*, 40, 553-564.

West, J. (2008). Commercializing open science: deep space communications as the lead market for Shannon Theory, 1960–73. *Journal of Management Studies*, 45, 1506–32.

Wong, P. K. and Singh, A. (2010). University patenting activities and their link to the quantity and quality of scientific publications, *Scientometrics*, 83(1), 271-294.

Wright, M., Lockett, A., Clarysse, B. and Binks, M. (2006). University spin-out companies and venture capital. *Research Policy*, 35, 481-501.

Zucker, L.G. and Darby, M.R. (2001). Capturing technological opportunity via Japan's star scientists: evidence from Japanese firms' biotech patents and products. *Journal of Technology Transfer*, 26, 37–58.

Zucker, L.G. and Darby, M.R. (2007). Virtuous circle in science and commerce. *Regional Science*, 86(3), 445-470.

Zucker, L.G., Darby, M.R. and Armstrong, J.S. (2002). Commercializing knowledge: university science, knowledge capture, and firm performance in biotechnology. *Management Science*, 48, 138–53.

Zucker, L.G., Darby, M.R. and Brewer, M. (1998). Intellectual human capital and the birth of US biotechnology enterprises. *American Economic Review*, 88(1), 290–305.